\documentclass[11pt,a4paper]{article}
\usepackage[hyperref]{acl2018}
\usepackage{times}
\usepackage{latexsym}
\usepackage{comment}
\usepackage{balance}
\usepackage{todonotes}

\usepackage{url}

\aclfinalcopy 

\title{Towards meaningful, grounded conversations with intelligent agents}

\author{
   Alexandros Papangelis \\
   Uber AI \\
   San Francisco, USA \\
   {\tt apapangelis@uber.com} \\\And
   Stefan Ultes \\
   Mercedes-Benz Research \& Development \\
   Sindelfingen, Germany \\
   {\tt stefan.ultes@daimler.com} \\
 }

\date{}

\begin{document}
\maketitle
\begin{abstract}
As conversational agents become integral parts of many aspects of our lives, current approaches are reaching bottlenecks of performance that require increasing amounts of data or increasingly powerful models. It is also becoming clear that such agents are here to stay and accompany us for long periods of time. If we are, therefore, to design agents that can deeply understand our world and evolve with it, we need to take a step back and revisit the trade-offs we have made in the current state of the art models. This paper argues that a) we need to shift from slot filling into a more realistic conversation paradigm; and b) that, to realize that paradigm, we need models that are able to handle concrete and abstract entities as well as evolving relations between them.
\end{abstract}

\section{Introduction}

Intelligent Conversational Agents (CA) 
have been around for several decades, primarily to provide information access or fulfil transactional needs. 
With the current unprecedented attention from the research community, applications such as robotic companions, personal assistants, customer support agents, or smart device agents are becoming more and more capable. Abstracting away from domain details, however, most of these agents are internally modelled by one of two broad approaches: {\bf transactional}, where the interaction revolves around achieving a task and dialogue structures such as dialogue states, actions, API calls, etc.\ are clearly defined and the conversation is grounded; and {\bf non-transactional}, where such structures are typically implicitly learned, with the goal of having a more casual conversation. 

Even though there have been many contributions to the state-of-the-art, the complexity of possible dialogue structures has remained rather limited. Instead, recent work proposed new algorithms \cite{su-EtAl:2017:SIGDIAL, casanueva2017benchmarking}, new state models \cite{schulz2017frame,lee-stent:2016:SIGDIAL}, or new system models \cite[among others]{wen2017,wenLIDM17,serban2016building,liu2017}. 

The goal of this paper is to argue that the complexity of dialogue structures is an important and relevant dimension of research. To address this, we propose a dialogue model that caters a high complexity while being suitable for all types of systems that operate with a symbolic state representation. We need to pursue research on the complexity of dialogue structures if we are to build real world CAs that combine the benefits of current works on transactional and non-transactional applications.

\section{Overview over Relevant Work} 

\begin{figure*}
    \centering
    \includegraphics[width=0.9\textwidth]{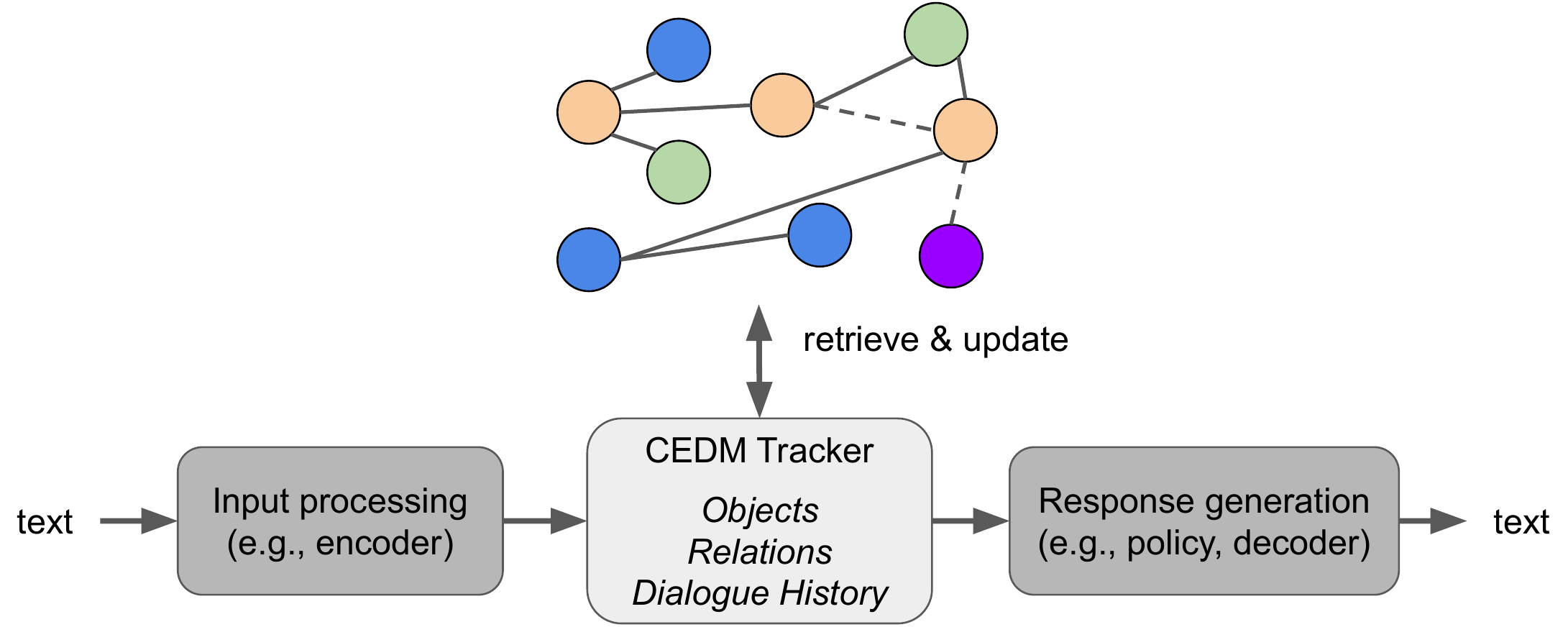}
    \caption{Overview of the proposed model: the CEDM-4-IN is at the centre of the system allowing input processing and response generation to be realised by any kind of model that operates with a symbolic state representation.}
    \label{fig:ced4in}
\end{figure*}  

To realise appropriate behaviour of a CA, machine learning algorithms have been used, especially reinforcement learning (RL) and deep learning (DL). Transformer-based approaches \cite{vaswani2017attention} have recently dominated the literature as they are powerful models able to capture relations between words in multiple levels, especially when pre-trained on very large corpora. In this section, we do not discuss works that only focus on parts of the dialogue pipeline, as depending on the downstream model and task they generally fall within the two broad approaches. Due to space constraints we discuss some representative works.

Transactional agents that build on machine learning have historically relied on RL to learn behaviours in a way that optimises a given objective which is specific to human-machine interaction rather than imitating human-human behaviour \cite[among others]{young2013,lemon2012,gasic2014gaussian,ultes2017pydial,vth-vu:2019:W19-59}. Due to various challenges, however, RL-based CAs are driven by dialogue models that limit the complexity of dialogue structures that can be processed by the agent. To remedy this, DL is being increasingly used for transactional and non-transactional agents alike, by having the CA learn to mimic human-human behaviour~\cite[e.g.]{serban2016building,li2016}. One way to combine DL with RL is to use DL as an approximator in an RL schema (Deep RL), for example \cite[]{wen2017,zhao-eskenazi-2016-towards,liu2017,papangelis2017single,su-EtAl:2017:SIGDIAL}. 

Recent work on transational CAs includes \citet{budzianowski2019hello}, who use a pre-trained GPT-2 \cite{radford2019language}, fine-tuned to a slot-filling task. They perform belief state tracking with the belief state being input to the network; however, no entity or relation tracking is performed. DialoGPT \cite{zhang2019dialogpt} is similar to GPT-2, but is trained on data closer to conversations (Reddit). It maximizes mutual information during training using a backwards model to predict the source sentence given the target sentence. These and some other modifications result in more ``conversational'' generated text. CAs such as the ones described above can work fairly well for restricted transactional applications, but often fail to provide substantive responses grounded to context, domain and world knowledge, or prior interactions with a particular user. To address these limitations, researchers have been attempting to incorporate domain or common sense knowledge. These approaches make the knowledge retrieval problem more explicit and force the model to learn not only how to respond but also how to retrieve relevant information.

\citet{eric-manning:2017:SIGDIAL} propose key-value retrieval networks, where an LSTM encoder learns to attend to the knowledge base (KB) and use that to condition the decoder. \citet{he2017learning} propose a model where each node in the KG is represented by an embedding. New nodes can be added and there is an attention mechanism to generate output. \citet{zhu2017flexible} propose GenDS, a system that can search for relevant entities and incorporate them in the output text, even if the entity is unseen during training. \citet{moon2019opendialkg} learn to traverse a KG by performing 1 or 2 hops starting from the current entity in order to inject interesting and relevant content. However, it is unclear what happens if the user mentions an entity further than 2 hops. \citet{fan2020augmenting} propose a KNN-based information fetching module that can combine multiple sources of information, such as similar dialogue contexts or general world knowledge. \citet{xu2020knowledge} use meta-learning to initialize the model's parameters, treat the graph as a whole, and use the entire conversation history as input. It is unclear, therefore, how this model can handle longer conversations or multiple conversations with the same user. \citet{xu2020dynamic} combine a dynamic knowledge graph and meta learning with adversarial samples (dynamic changes in the KG) for dialogue generation. 

Multi-modal CAs such as Robot CAs generally fall under the transactional category even if the task is not transactional as the added modalities and increased complexity (physical presence, gestures, etc.) require a more structured approach \cite[e.g.]{stiefelhagen2007,lucignano2013dialogue,marge2019research}.

Non-transactional CAs usually follow a model-free approach~\cite[e.g.]{serban2016building,li2016}. 
\citet{adiwardana2020towards} use an evolved transformer and performs better than other approaches (including DialoGPT). \citet{roller2020recipes} recently proposed a transformer-based non-transactional CA trained using various techniques (recipes) that result in natural-looking dialogues. Such models may perform well on a single turn or even a single dialogue. However, they would need large amounts of data originating from users who have multiple interactions with the agent over a long period of time to learn more complex behaviours, and even then they would probably learn the most frequent behaviours.

\section{Limitations of Current Works}

The works mentioned in the previous section have some shortcomings when it comes to modelling complex interactions, for example interactions with a CA that helps the user create art, analyse business data, or with an embodied CA that needs to have an understanding of its immediate surroundings and a deeper knowledge of the world as humans perceive it (dynamic objects, attributes, relations). In this paper, we advocate for a new conversational model combining two approaches \cite{ultes2018addressing,papangelis2018:SIGDIAL} that is powerful enough to handle conversations that are transactional, open, or that involve embodied agents in the physical world. Before presenting our approach, we discuss some of the main limitations of current state of the art CAs.

Approaches to transactional agents typically aim to tackle conversational coherence but not continuity, i.e., while they are successful in retrieving relevant knowledge for a single turn or even a single dialogue, it is hard to imagine how they can scale through time and handle multiple conversations with one user, remembering or forgetting information as needed. The bottleneck of current approaches (particularly end-to-end) is {\bf lack of longitudinal conversational data}, where users interact with the same agent multiple times over long periods. In current approaches, there is no notion of continuity or deeper understanding. It is up to the network to learn where to attend, how to retrieve and update its knowledge for every possible situation. What is more, none of the current works focuses on handling soft constraints and while a few can handle dynamic graphs (changes resulting from known relations) and unseen entities (usually unseen slot names or values), no work focuses on unseen relations. We summarise the main limitations of current works below:

\begin{description}
\item[Dialogue structure complexity.] Current CAs are limited in terms of dialogue structure complexity, in cases where content from a structured source needs to be incorporated.

\item[Entity modelling.] Entities and relations may be physical, conceptual, unseen, and dynamic.

\item[Lack of in-depth knowledge.] Lack of in-depth knowledge becomes obvious when the CA is sufficiently interrogated \cite{roller2020recipes}.

\item[Simple and repetitive language.] CAs show a tendency to stick to simpler language; and a tendency to repeat often-used phrases \cite{roller2020recipes}.

\item[Dialogue continuity.] Current end-to-end trained CAs usually need to re-do a lot of work at each time step (e.g., process input and history to resolve references, corrections, changes of user's mind).
\end{description}

\section{Conversational Entities for Information Navigation}

As discussed in the previous sections, there is a trend towards shifting the interaction paradigm from information retrieval (including question answering) to interactions involving large KGs that contain domain or world / common sense knowledge. However, current approaches still largely rely on big models such as transformers to capture patterns and nuances and learn when to do which API call or follow which KG path, in order to retrieve relevant knowledge. While this approach seems to work in some cases, it is hard to envision how it can scale with behaviour guarantees and interpretability (the ability to look at a user's logs and identify errors in the learnt behaviour, the API calls, or bugs in the KB, for example).

Information Navigation (IN) is a different paradigm of interaction that extends beyond information retrieval, question answering, or fetching a reasonable next utterance. IN focuses on having a conversation around several items or entities, comparing them, referring to their attributes, expressing hard and soft constraints over them, etc. This allows us to model applications from simple retrieval, information exploration (where the user has an idea of what they want and engage with the system to better understand their own needs) \cite{papangelis2017:SCAI, papangelis2018:SIGDIAL}, to open-ended discussions about concepts, to agents with physical presence that need to interact with and ground information on the physical world. An example implementation of IN is the Linked-Data SDS \cite{papangelis2017:SCAI, papangelis2018:SIGDIAL}, which is able to handle multiple operators between object attribute values ($<, \leq, >, \geq, =, \neq$, \emph{around, between, prefer} as well as combinations of these through logical operators $and, or, not$), multiple kinds of attributes (ordinal, nominal, hierarchical, set-valued, intervals) and soft and hard constraints. 

However, such CAs are still based on the flat domain-centred state which is not intuitive to model complex dialogue structures that link attributes of different objects (e.g.\ `I am looking for a hotel and a restaurant in the same area', `I need a taxi to the station in time to catch the train', `Bring the blue cup left of the vase.'), that represent multiple objects (e.g., two restaurants) or that connect to a set of objects (e.g., adding several participants to a calendar entry). The recently proposed Conversational Entity Dialogue Model (CEDM)~\cite{ultes2018addressing} aims at resolving these issues by moving away from the traditional domain-centred view towards an entity-centred approach. Each entity represents an object (of a certain type) or a relation between objects and allocates its own part of the dialogue state. This allows the dialogue system to easily maintain relations between objects, to process relations uttered by the user, and to address these relations in a system response.

A combination of these approaches will result in a dialogue model with all the properties necessary to process complex dialogue structures and to realize natural interaction: modelling of different operators between attributes and attribute values, modelling of relations between objects, introducing hard and soft constraints and being able to talk about each of these aspects. Further flexibility may be added by dynamically creating object type descriptions extracted from the KB thus relieving the system from the need for pre-defined object types. We therefore propose to combine CEDM and IN (CEDM-4-IN) in a more concentrated effort to shift the paradigm from information retrieval to information navigation. A paradigm where user and agent can have a conversation about entities that appear in the physical or conceptual world, understand their attributes and evolving relationships over longer periods of time. Figure~\ref{fig:ced4in} shows the proposed model, where the input processing and response generation parts may be any kind of models, from pipelines to deep neural networks.

\section{Addressing the limitations with CEDM-4-IN}

As discussed in the previous section, CEDM-4-IN is designed to model complex dialogue structures and therefore support complex information needs. CEDM also guarantees that the CA can have a deep understanding of the world and of the user's utterances, allowing it to make connections and inferences, provide interesting pieces of information, etc. Moreover, as our approach has a notion of state, it can more easily avoid repetitive language. Given a strong language generator, parts of the dialogue state can be used to condition it on the relevant history.

Perhaps the most interesting characteristic of CEDM-4-IN, however, is its ability to model objects and their relations (the dialogue state can be modelled as a KG) which can be permanent or transitory, physical or conceptual as shown in Table 1. 
This can open up a wide range of possibilities, given appropriately trained input processing and response generation models, allowing the CA to model long-term relationships (therefore dialogue continuity) by learning which kinds of entities or relations are permanent and which are transitory. This will enable CEDM-4-IN to better handle referring expressions, whether they concern events that happened in the past (e.g. `get the one from last time') or object characteristics (e.g. `the one with the red dots').

\section{Discussion}

\begin{table}
    \label{tab:object_examples}
    \centering
    \begin{tabular}{c|c|c}
        & Permanent & Transitory \\
        \hline
        Object & watch & meeting \\
        Relation & my pet & my cup \\
    \end{tabular}
    \caption{Examples of objects and relations: both can be permanent (`watch', `my' pet) or transitory (`meeting', `my' cup).}
\end{table}

So far we have shown how CEDM-4-IN can overcome some of the bottlenecks of current conversational models and push intelligent agents to a deeper understanding of the world. However, this expressive power does not come for free. Our approach requires dialogue state annotations that also cover entities and relations. It should be noted that this is the only kind of annotation that our model would need. Given such annotations, CEDM-4-IN will be able to generalise to unseen domains, as shown in \cite{papangelis2018:SIGDIAL}. In contrast, end-to-end models would require very large amounts of data consisting of multiple interactions for each user and will be tasked to learn many skills from scratch: conversation, information retrieval, knowledge update, etc. Such models usually have little guarantees and no interpretability. We argue, therefore, that CEDM-4-IN stands in a middle ground between traditional pipeline systems and end-to-end trainable approaches and is a more effective way to build next generation conversational agents.

\section{Conclusion and Outlook}

As conversational agents are becoming an integral part of our everyday lives, it is becoming increasingly evident that as a research field, we still have a long way to go. Aiming to shift the paradigm towards more realistic interactions with CAs over longer periods of time, we propose to combine two proven models, Information Navigation \cite{papangelis2018:SIGDIAL} and Conversational Entity Dialogue Model \cite{ultes2018addressing} into CEDM-4-IN. Our model can benefit from the interpretability and stronger behaviour guarantees that annotated data provide and also from the variability and generalizability that deep learning models offer. As future work we are planning to more concretely prove our approach on a benchmark conversational application.

\balance

\bibliographystyle{acl_natbib_new}
\bibliography{references}

\begin{thebibliography}{35}
\expandafter\ifx\csname natexlab\endcsname\relax\def\natexlab#1{#1}\fi

\bibitem[{Adiwardana et~al.(2020)Adiwardana, Luong, So, Hall, Fiedel,
  Thoppilan, Yang, Kulshreshtha, Nemade, Lu et~al.}]{adiwardana2020towards}
Daniel Adiwardana, Minh-Thang Luong, David~R So, Jamie Hall, Noah Fiedel, Romal
  Thoppilan, Zi~Yang, Apoorv Kulshreshtha, Gaurav Nemade, Yifeng Lu, et~al.
  2020.
\newblock Towards a human-like open-domain chatbot.
\newblock \emph{arXiv preprint arXiv:2001.09977}.

\bibitem[{Budzianowski and Vuli{\'c}(2019)}]{budzianowski2019hello}
Pawe{\l} Budzianowski and Ivan Vuli{\'c}. 2019.
\newblock Hello, it's gpt-2--how can i help you? towards the use of pretrained
  language models for task-oriented dialogue systems.
\newblock \emph{arXiv preprint arXiv:1907.05774}.

\bibitem[{Casanueva et~al.(2017)Casanueva, Budzianowski, Su, Mrk{\v{s}}i{\'c},
  Wen, Ultes, Rojas-Barahona, Young, and
  Ga{\v{s}}i{\'c}}]{casanueva2017benchmarking}
I{\~n}igo Casanueva, Pawe{\l} Budzianowski, Pei-Hao Su, Nikola
  Mrk{\v{s}}i{\'c}, Tsung-Hsien Wen, Stefan Ultes, Lina Rojas-Barahona, Steve
  Young, and Milica Ga{\v{s}}i{\'c}. 2017.
\newblock A benchmarking environment for reinforcement learning based task
  oriented dialogue management.
\newblock \emph{arXiv preprint arXiv:1711.11023}.

\bibitem[{Eric and Manning(2017)}]{eric-manning:2017:SIGDIAL}
Mihail Eric and Christopher~D. Manning. 2017.
\newblock \href {http://www.aclweb.org/anthology/W17-36 6} {Key-value retrieval
  networks for task-oriented dialogue}.
\newblock In \emph{Proceedings of the 18th Annual SIGdial Meeting on Discourse
  and Dialogue}, pages 37--49, Saarbrücken, Germany. Association for
  Computational Linguistics.

\bibitem[{Fan et~al.(2020)Fan, Gardent, Braud, and Bordes}]{fan2020augmenting}
Angela Fan, Claire Gardent, Chloe Braud, and Antoine Bordes. 2020.
\newblock Augmenting transformers with knn-based composite memory for dialogue.
\newblock \emph{arXiv preprint arXiv:2004.12744}.

\bibitem[{Ga{\v s}i{\'{c}} and Young(2014)}]{gasic2014gaussian}
Milica Ga{\v s}i{\'{c}} and Steve~J. Young. 2014.
\newblock {Gaussian processes for POMDP-based dialogue manager optimization}.
\newblock \emph{IEEE/ACM Transactions on Audio, Speech, and Language
  Processing}, 22(1):28--40.

\bibitem[{He et~al.(2017)He, Balakrishnan, Eric, and Liang}]{he2017learning}
He~He, Anusha Balakrishnan, Mihail Eric, and Percy Liang. 2017.
\newblock Learning symmetric collaborative dialogue agents with dynamic
  knowledge graph embeddings.
\newblock \emph{arXiv preprint arXiv:1704.07130}.

\bibitem[{Lee and Stent(2016)}]{lee-stent:2016:SIGDIAL}
Sungjin Lee and Amanda Stent. 2016.
\newblock Task lineages: Dialog state tracking for flexible interaction.
\newblock In \emph{SIGDial}, pages 11--21, Los Angeles. ACL.

\bibitem[{Lemon and Pietquin(2012)}]{lemon2012}
Oliver Lemon and Olivier Pietquin. 2012.
\newblock \href {https://doi.org/10.1007/978-1-4614-4803-7} {\emph{Data-Driven
  Methods for Adaptive Spoken Dialogue Systems}}.
\newblock Springer New York.

\bibitem[{Li et~al.(2016)Li, Monroe, Ritter, Galley, Gao, and
  Jurafsky}]{li2016}
Jiwei Li, Will Monroe, Alan Ritter, Michel Galley, Jianfeng Gao, and Dan
  Jurafsky. 2016.
\newblock Deep reinforcement learning for dialogue generation.
\newblock In \emph{Proceedings of the 2016 Conference on Empirical Methods in
  Natural Language Processing}, pages 1192--1202. Association for Computational
  Linguistics.

\bibitem[{Liu and Lane(2017)}]{liu2017}
Bing Liu and Ian Lane. 2017.
\newblock \href {https://doi.org/10.21437/Interspeech.2017-1326} {An end-to-end
  trainable neural network model with belief tracking for task-oriented
  dialog}.
\newblock In \emph{Proc. Interspeech 2017}, pages 2506--2510.

\bibitem[{Lucignano et~al.(2013)Lucignano, Cutugno, Rossi, and
  Finzi}]{lucignano2013dialogue}
Lorenzo Lucignano, Francesco Cutugno, Silvia Rossi, and Alberto Finzi. 2013.
\newblock A dialogue system for multimodal human-robot interaction.
\newblock In \emph{Proceedings of the 15th ACM on International conference on
  multimodal interaction}, pages 197--204.

\bibitem[{Marge et~al.(2019)Marge, Nogar, Hayes, Lukin, Bloecker, Holder, and
  Voss}]{marge2019research}
Matthew Marge, Stephen Nogar, Cory~J Hayes, Stephanie~M Lukin, Jesse Bloecker,
  Eric Holder, and Clare Voss. 2019.
\newblock A research platform for multi-robot dialogue with humans.
\newblock \emph{arXiv preprint arXiv:1910.05624}.

\bibitem[{Moon et~al.(2019)Moon, Shah, Kumar, and Subba}]{moon2019opendialkg}
Seungwhan Moon, Pararth Shah, Anuj Kumar, and Rajen Subba. 2019.
\newblock Opendialkg: Explainable conversational reasoning with attention-based
  walks over knowledge graphs.
\newblock In \emph{Proceedings of the 57th Annual Meeting of the Association
  for Computational Linguistics}, pages 845--854.

\bibitem[{Papangelis et~al.(2017)Papangelis, Papadakos, Kotti, Stylianou,
  Tzitzikas, and Plexousakis}]{papangelis2017:SCAI}
Alexandros Papangelis, Panagiotis Papadakos, Margarita Kotti, Yannis Stylianou,
  Yannis Tzitzikas, and Dimitris Plexousakis. 2017.
\newblock Ld-sds: Towards an expressive spoken dialogue system based on
  linked-data.
\newblock In \emph{Search Oriented Conversational AI, SCAI 17 Workshop
  (co-located with ICTIR 17)}.

\bibitem[{Papangelis et~al.(2018)Papangelis, Papadakos, Stylianou, and
  Tzitzikas}]{papangelis2018:SIGDIAL}
Alexandros Papangelis, Panagiotis Papadakos, Yannis Stylianou, and Yannis
  Tzitzikas. 2018.
\newblock Spoken dialogue for information navigation.
\newblock In \emph{Proceedings of the 19th Annual SIGdial Meeting on Discourse
  and Dialogue}, Melbourne, Australia. Association for Computational
  Linguistics.

\bibitem[{Papangelis and Stylianou(2017)}]{papangelis2017single}
Alexandros Papangelis and Yannis Stylianou. 2017.
\newblock Single-model multi-domain dialogue management with deep learning.
\newblock In \emph{International Workshop for Spoken Dialogue Systems}.

\bibitem[{Radford et~al.(2019)Radford, Wu, Child, Luan, Amodei, and
  Sutskever}]{radford2019language}
Alec Radford, Jeffrey Wu, Rewon Child, David Luan, Dario Amodei, and Ilya
  Sutskever. 2019.
\newblock Language models are unsupervised multitask learners.
\newblock \emph{OpenAI Blog}, 1(8):9.

\bibitem[{Roller et~al.(2020)Roller, Dinan, Goyal, Ju, Williamson, Liu, Xu,
  Ott, Shuster, Smith et~al.}]{roller2020recipes}
Stephen Roller, Emily Dinan, Naman Goyal, Da~Ju, Mary Williamson, Yinhan Liu,
  Jing Xu, Myle Ott, Kurt Shuster, Eric~M Smith, et~al. 2020.
\newblock Recipes for building an open-domain chatbot.
\newblock \emph{arXiv preprint arXiv:2004.13637}.

\bibitem[{Schulz et~al.(2017)Schulz, Zumer, Asri, and Sharma}]{schulz2017frame}
Hannes Schulz, Jeremie Zumer, Layla~El Asri, and Shikhar Sharma. 2017.
\newblock A frame tracking model for memory-enhanced dialogue systems.
\newblock \emph{arXiv preprint arXiv:1706.01690}.

\bibitem[{Serban et~al.(2016)Serban, Sordoni, Bengio, Courville, and
  Pineau}]{serban2016building}
Iulian~Vlad Serban, Alessandro Sordoni, Yoshua Bengio, Aaron~C Courville, and
  Joelle Pineau. 2016.
\newblock Building end-to-end dialogue systems using generative hierarchical
  neural network models.
\newblock In \emph{AAAI}, pages 3776--3784.

\bibitem[{{Stiefelhagen} et~al.(2007){Stiefelhagen}, {Ekenel}, {Fugen},
  {Gieselmann}, {Holzapfel}, {Kraft}, {Nickel}, {Voit}, and
  {Waibel}}]{stiefelhagen2007}
R.~{Stiefelhagen}, H.~K. {Ekenel}, C.~{Fugen}, P.~{Gieselmann}, H.~{Holzapfel},
  F.~{Kraft}, K.~{Nickel}, M.~{Voit}, and A.~{Waibel}. 2007.
\newblock Enabling multimodal human–robot interaction for the karlsruhe
  humanoid robot.
\newblock \emph{IEEE Transactions on Robotics}, 23(5):840--851.

\bibitem[{Su et~al.(2017)Su, Budzianowski, Ultes, Gasic, and
  Young}]{su-EtAl:2017:SIGDIAL}
Pei-Hao Su, Pawe{\l} Budzianowski, Stefan Ultes, Milica Gasic, and Steve Young.
  2017.
\newblock Sample-efficient actor-critic reinforcement learning with supervised
  data for dialogue management.
\newblock In \emph{SIGdial}, pages 147--157, Saarbrücken, Germany. ACL.

\bibitem[{Ultes et~al.(2018)Ultes, Budzianowski, Casanueva, Rojas-Barahona,
  Tseng, Wu, Young, and Ga{\v s}i{\' c}}]{ultes2018addressing}
Stefan Ultes, Pawe{\l~} Budzianowski, I\~nigo Casanueva, Lina Rojas-Barahona,
  Bo-Hsiang Tseng, Yen-Chen Wu, Steve Young, and Milica Ga{\v s}i{\' c}. 2018.
\newblock {Addressing Objects and Their Relations: The Conversational Entity
  Dialogue Model}.
\newblock In \emph{Proceedings of the 19th Annual SIGdial Meeting on Discourse
  and Dialogue}, Melbourne, Australia. Association for Computational
  Linguistics.

\bibitem[{Ultes et~al.(2017)Ultes, Rojas-Barahona, Su, Vandyke, Kim, Casanueva,
  Budzianowski, Mrk{\v s}i{\'{c}}, Wen, Ga{\v s}i{\'{c}}, and
  Young}]{ultes2017pydial}
Stefan Ultes, Lina~M. Rojas-Barahona, Pei-Hao Su, David Vandyke, Dongho Kim,
  I{\~{n}}igo Casanueva, Pawe{\l} Budzianowski, Nikola Mrk{\v s}i{\'{c}},
  Tsung-Hsien Wen, Milica Ga{\v s}i{\'{c}}, and Steve~J. Young. 2017.
\newblock Pydial: A multi-domain statistical dialogue system toolkit.
\newblock In \emph{ACL Demo}. Association of Computational Linguistics.

\bibitem[{Vaswani et~al.(2017)Vaswani, Shazeer, Parmar, Uszkoreit, Jones,
  Gomez, Kaiser, and Polosukhin}]{vaswani2017attention}
Ashish Vaswani, Noam Shazeer, Niki Parmar, Jakob Uszkoreit, Llion Jones,
  Aidan~N Gomez, {\L}ukasz Kaiser, and Illia Polosukhin. 2017.
\newblock Attention is all you need.
\newblock In \emph{Advances in neural information processing systems}, pages
  5998--6008.

\bibitem[{Väth and Vu(2019)}]{vth-vu:2019:W19-59}
Dirk Väth and Ngoc~Thang Vu. 2019.
\newblock \href {https://www.aclweb.org/anthology/W19-5908} {To combine or not
  to combine? a rainbow deep reinforcement learning agent for dialog policies}.
\newblock In \emph{Proceedings of the 20th Annual SIGdial Meeting on Discourse
  and Dialogue}, pages 62--67, Stockholm, Sweden. Association for Computational
  Linguistics.

\bibitem[{Wen et~al.(2017{\natexlab{a}})Wen, Miao, Blunsom, and
  Young}]{wenLIDM17}
Tsung-Hsien Wen, Yishu Miao, Phil Blunsom, and Steve Young. 2017{\natexlab{a}}.
\newblock Latent intention dialogue models.
\newblock In \emph{ICML}, ICML'17. JMLR.org.

\bibitem[{Wen et~al.(2017{\natexlab{b}})Wen, Vandyke, Mrk{\v s}i{\'{c}}, Ga{\v
  s}i{\'{c}}, Rojas-Barahona, Su, Ultes, and Young}]{wen2017}
Tsung-Hsien Wen, David Vandyke, Nikola Mrk{\v s}i{\'{c}}, Milica Ga{\v
  s}i{\'{c}}, Lina Rojas-Barahona, Pei-Hao Su, Stefan Ultes, and Steve Young.
  2017{\natexlab{b}}.
\newblock A network-based end-to-end trainable task-oriented dialogue system.
\newblock In \emph{EACL}, pages 438--449. ACL.

\bibitem[{Xu et~al.(2020{\natexlab{a}})Xu, Bao, and Wang}]{xu2020knowledge}
Hongcai Xu, Junpeng Bao, and Junqing Wang. 2020{\natexlab{a}}.
\newblock Knowledge-graph based proactive dialogue generation with improved
  meta-learning.
\newblock \emph{arXiv preprint arXiv:2004.08798}.

\bibitem[{Xu et~al.(2020{\natexlab{b}})Xu, Bao, and Zhang}]{xu2020dynamic}
Hongcai Xu, Junpeng Bao, and Gaojie Zhang. 2020{\natexlab{b}}.
\newblock Dynamic knowledge graph-based dialogue generation with improved
  adversarial meta-learning.
\newblock \emph{arXiv preprint arXiv:2004.08833}.

\bibitem[{Young et~al.(2013)Young, Ga{\v s}i{\'{c}}, Thomson, and
  Williams}]{young2013}
Steve~J. Young, Milica Ga{\v s}i{\'{c}}, Blaise Thomson, and Jason~D. Williams.
  2013.
\newblock {POMDP-based statistical spoken dialog systems: A review}.
\newblock \emph{Proceedings of the IEEE}, 101(5):1160--1179.

\bibitem[{Zhang et~al.(2019)Zhang, Sun, Galley, Chen, Brockett, Gao, Gao, Liu,
  and Dolan}]{zhang2019dialogpt}
Yizhe Zhang, Siqi Sun, Michel Galley, Yen-Chun Chen, Chris Brockett, Xiang Gao,
  Jianfeng Gao, Jingjing Liu, and Bill Dolan. 2019.
\newblock Dialogpt: Large-scale generative pre-training for conversational
  response generation.
\newblock \emph{arXiv preprint arXiv:1911.00536}.

\bibitem[{Zhao and Eskenazi(2016)}]{zhao-eskenazi-2016-towards}
Tiancheng Zhao and Maxine Eskenazi. 2016.
\newblock \href {https://doi.org/10.18653/v1/W16-3601} {Towards end-to-end
  learning for dialog state tracking and management using deep reinforcement
  learning}.
\newblock In \emph{Proceedings of the 17th Annual Meeting of the Special
  Interest Group on Discourse and Dialogue}, pages 1--10, Los Angeles.
  Association for Computational Linguistics.

\bibitem[{Zhu et~al.(2017)Zhu, Mo, Zhang, Zhu, Peng, and
  Yang}]{zhu2017flexible}
Wenya Zhu, Kaixiang Mo, Yu~Zhang, Zhangbin Zhu, Xuezheng Peng, and Qiang Yang.
  2017.
\newblock Flexible end-to-end dialogue system for knowledge grounded
  conversation.
\newblock \emph{arXiv preprint arXiv:1709.04264}.

\end{thebibliography}

\end{document}